# New class of biological detectors for WIMPs.

A.K. Drukier [1,2], Ch. Cantor [3], M. Chonofsky [4], G.M. Church [5], R.L. Fagaly [6] K. Freese [7], A. Lopez[7], T. Sano [8], C. Savage[9], W.P. Wong [4]

(1):  BioTraces Inc., 5660 Oak Tanger Ct, Burke, VA. 22015, USA(adrukier@gmail.com)
(2)   Physics Department, George Mason University, Fairfax, VA., USA
(3)   Sequenom Inc., John Hopkins Ct., San Diego, Ca. 92121 (ccantor@sequenom.com)
(4)   Departments of Biological Chemistry and Molecular Pharmacology and Pediatrics, Harvard Medical School and Boston Children's Hospital (wesley.wong@childrens.harvard.edu)
(5)   Department of Genetics, 77 Avenue Louis Pasteur, Harvard Medical School, Boston MA 02115 USA (gchurch@harvard.edu)
(6)   BioTraces Inc., i*bidem*, (robert.fagaly@gmail.com)
(7)   Michigan Center for Theoretical Physics, Department of Physics, University of Michigan, Ann Arbor, MI 48109,USA (ktfreese@umich.edu and aolopez@umich.edu)
*(8)  DiThe*ra, Inc., San Diego, CA 92121 (sanotakeshi@gmail.com)
*(9)* Department of Physics & Astronomy, University of Utah, Salt Lake City, UT (savage@physics.utah.edu)

December 2013

**Abstract:**   Weakly Interacting Massive Particles (WIMPs) may constitute a large fraction of the matter in the Universe. There are excess events in the data of DAMA/LIBRA, CoGeNT, CRESST-II, and recently CDMS-Si, which could be consistent with WIMP masses of approximately 10 GeV/$c^2$. However, for $M_{DM} > 10$ GeV/$c^2$ null results of the CDMS-Ge, XENON, and LUX detectors may be in tension with the potential detections for certain dark matter scenarios and assuming a certain light response.

 We propose the use of a new class of biological dark matter (DM) detectors to further examine this light dark matter hypothesis, taking advantage of new signatures with low atomic number targets, Two types of biological DM detectors are discussed here: DNA-based detectors and enzymatic reactions (ER) based detectors. In the case of DNA-based detectors, we discuss a new implementation.  In the case of  ER detectors, there are four crucial phases of the detection  process:
a)   change of state due to energy deposited by a particle;
b)   amplification due to the release of energy derived from the action of an enzyme on its substrate;
c)   sustainable but non-explosive enzymatic reaction;
d)   self-termination due to the denaturation of the enzyme, when the temperature is raised.
This paper provides information of how to design as well as optimize these four processes.

## 1) INTRODUCTION

In the mid 20$^{th}$ century, new classes of particle/radiation detectors were introduced about every 5 years. The last new classes of detectors were introduced about 30 and 25 years ago, namely cryogenic particle detectors and liquid noble gas detectors, respectively. Yet there are needs of physics and biology which require new classes of detectors with nano-metric spatial resolution *e.g.* detection of Dark Matter (DM) candidates and detectors for mass-spectroscopy.

The first generation of particle/radiation detectors consisted of photographic emulsions and gas detectors.  The majority of modern particle detectors are liquids and solid-state detectors. All of these detectors have, as output, either photons or electrons, which are easy to count with modern photonics and/or electronics.

We propose a new class of detectors, which use thermal processes or molecular transformations to detect particle interaction effect(s). One of the byproducts of the development of molecular biology and nanotechnology is that we now understand and can engineer material properties at a scale of a few nanometers. This includes better understanding of heat propagation processes at the nanoscale. We can produce a wide variety of nano-size objects and order them spatially.  It becomes possible to manipulate the flow of heat just as we manipulate the flow of electrons in solid-state devices. In a sense, the development of Superheated Superconducting Colloid [26] and other cryogenic bolometers [18, 19, 24] were a first step in this direction. However, they operate at cryogenic temperatures, which facilitate thermal engineering but make their implementation and operation more difficult. In this paper, we propose bio-detectors that can operate as nano-size bolometers at room temperature. Optimization of DNA-based and enzymatic reactions (ER) particle detectors is a new, multifaceted field. We describe the principles on which these detectors are based and briefly describe the particular design steps.

**ssDNA-based detectors:** One class of such detectors is single-stranded DNA (ssDNA) based detectors. It was described by A.K. Drukier *et al*. (2012) and is illustrated in Figures 1 and 2 [32]. The acquisition stage is followed by a development stage, *i.e.,* ssDNA-based detectors are not real-time detectors. The energy, which has been deposed by recoiling nuclei, is used to break ssDNA molecules that are long biological polymers and serve as detector elements. Then, the broken fragments of DNA strands are counted, and their lengths are measured. Use of a high-density array of ssDNA as a detector could offer spatial resolution of the order of a few nanometers [32]. Recent simulations suggest that the diurnal modulation effect (DME) may be observed in such detectors.

In Section II, we demonstrate the need for low atomic number detectors for reliable detection of DM candidates, particularly when they have low masses ($M_{DM} < 10$ GeV/c$^2$). Then we introduce a ssDNA detector and describe some simulations that suggest their ability to detect the direction of incoming DM particles. In the third part of this paper, we describe biological detectors based on "enzymatic reactions".

**The enzymatic reaction (ER) detectors**: Elsewhere some of us proposed nanoexplosives as dark matter detectors [25]. When the dark matter hits the detector, it warms it up enough to cause an explosion, first in a small nano-sized region and then by chain reaction of a larger region. Here an alternative is proposed that is quite similar but more controlled. Enzymatic reactions (ER) can cause a more limited runaway. The reaction takes place only in a small temperature range: when the temperature is either too small or too large, the ER stops. If the temperature is too low, the diffusion of the substrate towards the enzyme does not take place; if the temperature is too high, the enzyme is denatured. Unlike explosions, which continue as long as the temperature is high, the ER reactions are more controlled.

The ER detector is a cubic meter "block of "ice" on surface of which is placed an array, say 64, of microphones. The ice consists of a mixture of enzyme (< 1%), substrate, e.g. {(10%) $H_2O_2$ + 90% $H_2O$}., and Li nano-grains. As long as the mixture remains frozen, no reaction takes place. When the DM interacts with this mixture, WIMP scatters with the Li, deposits energy into the mixture, and melts some of the ice. A 100-1,000 nm$^3$ volume of water is created, and the enzyme can now come into contact with the substrate and an enzymatic reaction is initiated, e.g., $2H_2O_2$ + catalase $\rightarrow$ $2H_2O$ + $O_2$ + energy. The reaction leads to a few micron size bubble of gas, the ice cracks and sound results. The sonic data are analyzed to localize the x-y-z position and total energy in the event.

In the proposed applications, the enzymatic reaction must be triggered or sensitized. Enzymes do not work in ice. The substrate to be acted on is transported towards enzyme by diffusion, *ergo* freezing => no diffusion => no enzymatic reaction. One may trigger the enzymatic reaction by melting the substrate. We describe an ER-detector which is a frozen mixture of substrate and enzyme, which is melted and triggered only at or close-to the track of the particle. The substrate will consist of mixture of $H_2O$ and $D_2O$ seeded with nano-grains of low mass nuclei, *e.g.* Li, Be, or B. It is possible to built the hierarchical substrate consisting of nano-grains {$R_{grain}$ =O(10 nm)} of substrate/enzyme ice, dense packed into nano-balls {$R_{ball}$=O(500 nm)} and then ordered into foils or slabs. By thermal engineering one can stop the avalanche propagation at the border of a single nano-ball. All these techniques are well developed by nanotechnology but have never previously been applied to enzymatic reactions, even if a similar structures exists in biology, *e.g.* vesicles filled with biological materials, including a particular enzyme or a mixture of enzymes.

The ER-detectors, are based on both thermodynamics applied to nano-objects and molecular properties of enzymes acting on particular substrates. In electronics, the crucial role play processes in which electronic signals are triggered and amplified. In ER detectors, the trigger is change of state, *i.e.,* melting of ice very close to recoiling nuclei trajectory or at the vertex of an interaction. Here, the use of the smallest objects that are compatible with a notion of temperature is preferred. This means that size of the objects must be larger than, for example, 1,000 atoms or a few nanometers in size. For such objects, the specific heat at room temperature (RT) is very small indeed; $C_v$ =

$O(10^{-5}$ keV/nm$^3)$. Even if the thermal processes are orders of magnitude slower than electronic processes, at a few nanometer scale, they are fast enough for most applications. Also, the equivalent of electronic resistance is well defined; especially if granular objects are placed in a sufficiently different material medium, *e.g.*, metal grains in plastic or liquid in a shell. There is an efficient phonon mismatch; *i.e.*, small grains are thermally insulated. The time scale of thermalization inside a grain can be made much shorter than the time scale of heat escape. Thus, we can treat a collection of nano-grains made of ice as a collection of independent heat sensitive elements and can reliably calculate if they will change state when a particular amount of heat is deposited by particle interaction with a lattice of such ice nano-grains. Note the similarity with a CCD—heat is first stored in some geometrically well-defined structures and then escape leads to predictable effects.

A change of state in a single grain with a radius of a few nanometers is very difficult to detect. However, we can amplify it. by using nano-explosives and/or enzymatic reactions. We note that the amount of energy density stored in the best electric battery is about 100-fold smaller than that in combustables. However, a fuel + oxygen combination has orders of magnitude lower energy density than a high explosive. In a companion paper, we present a nano-explosives detector, in which one can with 0.1 keV ignite a $V = 1,000$ nm$^3$ grain, which leads to the release of about 5 MeV of thermal energy. Thus, in a $V = 1,000$ nm$^3$ high explosive grain, it is possible to achieve a (5 MeV/0.1 keV) = $5 \times 10^4$ energy amplification. Let's compare this with a typical CMOS amplifier; here the amplification of 100 is achieved in a structure with $V = 64$ nm x 64 nm x 10 nm. In the language of electrical engineering we would call a single nano-explosive grain a low threshold but high gain circuit. By using nano-explosives, we can achieve an amplification per unit volume, which is about 100-fold better than in the best electronic amplifiers. The challenge of nano-explosive detectors is not how to trigger the explosive energy release, but how to control it and localize it in, for example, a 1 μm$^3$ volume. The density of energy that is derived from the most efficient enzymatic reactions is smaller, but a 1,000-fold amplification seems possible. In this paper, we suggest that it is easier when an enzymatic reactions with high, but lower than high explosive, energy content are used.

In ER detectors, there are four crucial phases of the proposed process:

a)  change of state due to energy deposited by a particle;
b)  amplification due to the release of energy derived from the action of an enzyme on its substrate;
c)  sustainable but non-explosive enzymatic reaction;
d)  self-termination due to the denaturation of the enzyme, when the temperature is raised to 40-80$^o$C (depending on enzyme).

This paper provides some information of how to design as well as to optimize these four processes.

There are hundreds of enzymes, which are potentially effective in this application. The majority of them work at temperatures close to room temperature (RT) which is very convenient for detector applications. Some of them work at temperatures as low as 5-10 °C, *i.e.* close to the temperature of melting ice. A few work at temperatures below 0 $^o$C, not in water but in low freezing temperature mixtures. In the following, we will focus on catalase, which is a well studied, very efficient and yet relatively low cost enzyme. The enzyme itself is always < 1 % per mass, and enzymatic reactions takes place in an aqueous medium. Thus, we always have $\rho = O(1$ g/cc$)$. However, the said medium can be seeded with a high density material in which the recoil nuclei are adsorbed.

**The vertex detector concept:** ER-detectors will work best if the energy is deposed in a single "voxel". When neutral particles scatter on nuclei; *e.g.*, the case of DM particles, fast neutrons, or neutrinos; the majority of the energy is transferred to the lattice and recoiling nuclei. It leads to the creation of ballistic phonons, which rapidly thermalize and increase grain temperature. For low-mass DM particles and their expected velocity, the energy of the recoiling nuclei is 0.1-5 keV. All of this energy is deposited within a few nanometers. Thus, the dE/dx = $O(1$ keV/nm$)$ is deposited in the vertex, *i.e.* in grain in which a DM candidate interacts. This leads to the propagation of spherical heat wave, which is simple to understand, resulting in the change of state.

The energy deposition is much smaller in the case of single charged, relativistic particles, which have a range of hundred of μm in materials with $\rho = O(1$ g/cc$)$. This corresponds to dE/dx < 1 eV/nm. Even for alpha particles, dE/dx is < 10 eV/nm. We have to compare the energy deposited to the energy necessary to change the state, which is proportional to the volume of the grain. The large difference in dE/dx means that the detection of single charged particles and alpha particles is highly suppressed.

## 2) THE CHALLENGE: DETECTING DM

The Milky Way, along with other galaxies, is encompassed in a massive dark matter (DM) halo of unknown composition[12] . Only about 5% of the Universe is ordinary baryonic matter, while the remainder is about 25% dark matter and 70% dark energy [3, 4]. Identifying the nature of the dark matter is a long standing problem [1]. Leading candidates for the DM are Weakly Interacting Massive Particles (WIMPs), a generic class of particles that includes the lightest supersymmetric (SUSY) particle[14]. These particles were produced in the early universe and annihilated with one another so that a predictable number of them remain today.  Thermal WIMPs are good DM candidates because, assuming only weak interactions for the particle annihilation, *ad novo* calculations are compatible with DM density as estimated from WMAP and Planck data to be about 25%.  Furthermore, Asymmetric Dark Matter candidates are quite attractive [15].

In the mid-1980s, a series of seminal and synergistic papers [5-9] created a new field of direct searches for DM. It also established a "WIMP orthodoxy", which is based on the following statements:
2.1) Only about 5% of matter in the Universe consists of ordinary atomic matter, and the rest is unknown (Dark Matter/Energy);
2.2) WIMPs, including those predicted by SUSY, are plausible candidates for DM;
2.3) WIMP candidates are expected to have masses of 1 GeV – 10 TeV;
2.4) WIMPs interact with nuclei, and, for WIMP velocities at the peak of the distribution *i.e.* O(200) km/sec in the galactic frame, transfer up to a few tens keV energy*; i.e.,* the resulting recoil nuclei are non-relativistic and couple weakly with electrons; thus, the "best detectors" are bolometers and not ionization detectors;
2.5) The main challenge is not detection but the signal/noise ratio.

For a SUSY neutralino and many other WIMP candidates, the dominant WIMP-quark couplings in direct detection experiments are the scalar and axial-vector couplings, which respectively give rise to spin-independent (SI) and spin-dependent (SD) cross-sections. Following the "WIMP orthodoxy", almost all experiments focus on the case of SI interactions, in which the total SI cross-section of a nucleus is a coherent sum over the individual protons and neutrons within and scales as $A^2$, *i.e.* with the square of the atomic mass.  In addition, for WIMPs that are much heavier than the nucleus, due to the kinematic factors that also scale as $A^2$, the cross-section scales as $A^4$. However, the form factor suppression becomes more significant as the size of the nucleus increases, so the scattering rate does not scale as $A^4$ for very heavy nuclei, though it still rises rapidly with A (at least as fast as $A^2$).  As a result, direct detection experiments often use heavy nuclei to increase their sensitivity to WIMP scattering.  However, low mass WIMPs are inefficient at transferring their kinetic energy to heavy nuclei, leading to relatively low recoil energies.  While interaction rates with light WIMPs are still maximized by heavy nuclei, the low energies of those events means that they will mostly go undetected.  As the nuclear recoil energies are maximized when the nuclear mass is comparable to the WIMP mass, the count rate for *detectable* events in the case of light WIMPs will be maximized for nuclei not much heavier than the WIMP (though the ideal nuclear target will depend on the energy threshold of the detector).

A multitude of experimental efforts are currently underway to detect WIMPs, with some claiming detection. The count rate in direct detection experiments experiences an annual modulation [2, 11-12] due to the motion of the Earth around the Sun. Because the relative velocity of the detector with respect to the WIMPs depends on the time of year, the count rate exhibits (in most cases) a sinusoidal dependence with time.  For the standard dark matter Halo model, the flux is maximal in June and minimal in December.  Annual modulation is a powerful signature for dark matter because most background signals, *e.g.* from radioactivity in the surroundings, are not expected to exhibit this kind of time dependence. It would be even better if one could detect the direction of the incoming DM particle as the direction of the incoming WIMPs are strongly peaked towards the direction of the Sun's motion about the center of the Galaxy. New signatures are possible when the detectors can recognize the track of recoiling nuclei with a precision of a few nanometers .

For more than a decade, the DAMA experiment using NaI crystals [16] has been detecting an annual modulation.

The DAMA annual modulation is currently reported as almost a 9 sigma effect [16] , and is consistent with either about 10 GeV/c$^2$ or 80 GeV/c$^2$ , *i.e.* due to scattering on Na or I, respectively [16, 23] . The CoGeNT experiment reported a 2.8σ evidence for an annual modulation [17]. Additionally, CRESST-II and CDMS-Si have also announced excess over background [18,19]. Whether DAMA, CoGeNT, and CRESST are consistent in the low WIMP mass O(10 GeV/c$^2$) window is still under debate [23-25]. Yet CDMS-Ge [18], SuperCDMS [18], XENON [20,21] and LUX [22] find null results, that appear to be in conflict with the four experiments that report excess counts over background.

In dark matter experiments, signal to background (S/B) is always poor, particularly for low-mass DM. Thus, we need SIGNATURES that can be used to validate that count we observe are DM candidates. The most reliable signatures are: the annual modulation effect and other directional effects, an appropriate dependence of count rate on atomic mass (*e.g.* A$^2$), reliable information that the range of the particle(s) is below 100 nm, and a particular ratio of FM = (total energy deposed)/(energy transferred to electrons).

*Kinematics:* WIMP direct detection experiments seek to measure the energy deposited when a WIMP interacts with a nucleus in a detector. If a WIMP of mass $M_{DM}$ scatters elastically from a nucleus of mass $M_n$, it will deposit a recoil energy:

$$Enr = (\mu^2 V^2 / M_n)(1 - \cos(\theta)),$$

where $\mu = M_{DM} M_n / (M_{DM} + M_n)$ is the reduced mass of the WIMP-nucleus system, V is the speed of the WIMP relative to the nucleus, and θ is the scattering angle in the center of mass frame. The differential recoil rate per unit detector mass, typically given in units of cpd kg$^{-1}$ keV$^{-1}$ (where cpd is counts per day), can be written as:

$$\frac{dR}{dE} = \frac{n_{DM}}{M_n} \left\langle V \frac{d\sigma}{dE} \right\rangle = \frac{1}{2 M_n \mu^2} \sigma(q) \rho_{DM} \eta(v_{min}(E_{nr}, t)),$$

where $n_{DM} = \rho_{DM} / M_{DM}$ is the number density of WIMPs, with $\rho_{DM}$ the local dark matter mass density; $q = \sqrt{2 M_n E_{nr}}$ is the momentum exchange in the scatter; $\sigma(q)$ is an effective scattering cross-section;

$$\eta(V_{min}, t) = \int_{V > V_{min}} d^3 V \frac{f(\vec{V}, t)}{V}$$

is the mean inverse velocity with $f(\vec{V}, t)$ the (time-dependent) WIMP velocity distribution; and $V_{min} = \sqrt{E_{nr} M_n / (2 \mu^2)}$ is the minimum WIMP velocity that can result in a recoil energy $E_{nr}$. The typical energy transferred to the nucleus in a scattering event is from 1 to 50 keV, depending on $M_{DM}$ and velocity. Typical count rates in detectors are less than 1 count per kg of detector per day.

Over the past twenty five years a variety of designs have been developed to detect WIMPs. They include detectors that measure scintillation; ionization; and dilution-refrigerator based calorimeters which measure the total energy deposed by means of a phonon spectrum. Current detector masses range in size up to 100 kg (e.g. DAMA, XENON-100, LUX). The plan for the next generation of detectors is to reach one tonne.

A major concern in all WIMP detectors is backgrounds. To eliminate spurious events from CR, the detectors must be placed deep underground ( > 2,000 m of water equivalent). Yet some radioactive backgrounds remain and must be eliminated. Thus the experimental determination of annual and/or diurnal modulation is a crucial test of the WIMP origin of any events observed in the detector, as most backgrounds should not exhibit the same time dependence.

With $M_{DM}$ < 10 Gev/c$^2$, the main challenge is to provide a detector with A = O(10) and $E_{th}$ < 1 keV. There is a need for: effective transfer of energy from a DM particle to a target nucleus, *i.e.,* $M_n$ = O($M_{DM}$); high energy loss dE/dx, which is a very strong function of the material; and very efficient background rejection, which is possible when the range of the particle is known.

The maximum energy transferred by a WIMP is for the case theta = 180 degrees, so that $E_{max} = 2\mu^2 V^2 / M_n$.. For an optimal match, $M_n = M_{DM}$, we have $\mu \Rightarrow \frac{1}{2} M_{DM}$ and $E_{max} \Rightarrow \frac{1}{2} M_{DM} V^2$. For a reasonable match $1/3 < M_A/M_{DM} < 3$. Importantly, for a detector with given threshold $E_{th}$ we have the limit on the velocity of detected DM, namely:

$$E_{max} > E_{th} \Rightarrow V > V_{min,th} = 212.5 \sqrt{[(E_{th} \cdot M_n)/m]}$$

wherein $V_{min,th}$ is in km/sec, $E_{th}$ is in keV, $M_n$ and $M_{WIMP}$ in GeV/c$^2$.

One needs $V_{min,th} << V_{cutoff}$, wherein $V_{cutoff}$ is the cutoff in the velocity distribution in the Earth frame resulting from the fact that WIMPs in the galactic rest frame moving faster than escape velocity do not stay in the Galaxy. We take the escape velocity in the galactic rest frame to have a value of about 550 km/sec for the most popular model of DM halo in a galaxy. Converting to the Earth frame of the detector, this corresponds to $V_{cutoff}$ roughly 750 km/sec. The galactic models are quite uncertain close to $V_{escape}$, and any conclusions, drawn from the detection of DM moving in the galactic frame with velocities in excess of ~500 km/sec, are uncertain.

Elsewhere [25] some of us calculated $V_{min}$ for 15 different detectors. Among the said detectors are: (1) = Be-ssDNA, (2) = Enzymatic-reaction detectors, (3) = Be-SSC, (4) = DAMA, (5) = CoGeNT, (6) = CDMS-Si, (7) = CRESST, (8) = CDMS-Ge and (9) = liquid Xenon. Detectors (1) and (2) are discussed in this paper. We demonstrated that only detectors (1), (2) and (3), as well as nano-boom detectors (described in a companion paper in this collection), have $V_{min} < V_{peak}$ for $M_{DM} = O(5 \text{ GeV/c}^2)$. $V_{peak}$ is the velocity at the peak of distribution, *i.e.* $V_{peak} \approx 300$ km/sec in the Earth's frame.

Most existing detectors are made of high mass materials that are not ideal for the study of low mass WIMPs. For example, for 5 GeV/c$^2$ WIMPs, liquid xenon detectors with an energy threshold of 4.3 keV have $V_{min\_} = 1016$ km/sec, and with an energy threshold of 3 keV have $V_{min\_} = 850$ km/sec. However, there are essentially no WIMPs with these velocities coming into the detectors (due to cutoff of WIMPS at the escape velocity from the Galaxy).

To maximize the energy transferred, we need deuterium or helium based detectors if $M_{DM} = O(3 \text{ GeV/c}^2)$ and Li, Be, or B based detectors if $M_{DM} = O(5 \text{ GeV/c}^2)$. The mixed material may be used to maximize both the energy transfer (low A) and stopping power for recoiling nuclei (high Z, high density). Interaction on (C, N, O) are marginally acceptable for $M_{DM} = O(5 \text{ GeV/c}^2)$ if $E_{th} < 1$ keV. Importantly, even if cross-sections of the interaction with DM candidates fall fast, because low A materials need to be used; it is not signal, but signal to background (S/B), which is limiting.

**Directional Detectors and the Diurnal Modulation Effect:** A major step forward in the field of direct detection would be the development of detectors with directional capability [12, 13], i.e., the capability to determine which direction the WIMP came from. As a result of the elastic scattering of a WIMP off of a nucleus in the detector, the nucleus gets kicked in a particular direction (typically forward).
Thus by determining the track of the nucleus one could identify the direction of the incoming WIMP. The WIMP flux in the lab frame is peaked in the direction of motion of the Sun (which happens to be towards the constellation Cygnus). Hence the recoil spectrum for most energies should be peaked in the direction opposite to this. The event rate in the backward direction is expected to be ~10 times larger than that in the forward direction [12,35]. If a detector has a non-uniform sensitivity to recoils in different directions (which is not typical as most detectors are insensitive to the recoil direction), that detector will experience a diurnal modulation in the observed event rate due to the changing orientation of the detector relative to the incoming WIMP wind as the Earth rotates. For example, the ssDNA detector described below uses a thin foil target, with only nuclei recoiling out of one side of the foil being detected (it is sensitive to only $2\pi$ of the full $4\pi$ of angular parameter space). The statistical requirements to identify this diurnal modulation would only require ~ 30-100 WIMP interactions [31]. Some experiments, such as DRIFT [36], aim to reconstruct individual recoil tracks, which allows for the changing incoming WIMP wind direction (relative to the lab frame) to be explicitly observed and could provide further information regarding the distribution of WIMPs in the halo. Measurements of these diurnal modulations could then provide –as for the case of annual modulation – a "smoking gun" for the detection of WIMPs. In addition, any galactic substructure in the WIMP density, such as tidal streams, could show up as spikes coming from one particular direction in a directional detector, but for this effect to be measurable one would need a few tonne detector.

**Energy Loss of Recoling Nuclei:** When a WIMP hits a nucleus in the detector, the recoiling nucleus is propelled forward into the detector medium with energy in the O(1-10)keV range. We need to know the range of the recoiling nucleus in the detector; i.e. how far it goes before it stops. Thus we compute the typical energy loss per unit length due to interactions of the nucleus with the detector. Particles passing through a bulk of material interact with the nuclei and electrons of the media and consequently lose energy. The stopping power, which measures the energy loss per length, has been a source of much interest due to its applicability in material science and medicine. The stopping power depends on the energy, particle and media considered. There have been multiple equations formulated in order to predict analytically the stopping power. The Lindhard equation is used to calculate dE/dx for a slowly moving heavy ion moving through a medium (typical energies for the ion are a few keV):

$\frac{dE}{dx} = NS_{total} = N(S_n + S_e)$, where N is the number density of the medium, $S_{total}$ is the total stopping power, $S_n$ is the nuclear stopping power, and $S_e$ is the electronic stopping power. For low energy ions, in which the ion energy is less than a few keV, the nuclear stopping power dominates over the electronic stopping power and is given by:

$$S_n = \frac{C_m E^{1-2m}}{1-m} \left[ \frac{4 M_1 M_2}{(M_1 + M_2)^2} \right]^{1-m},$$

where E is the energy (eV), $M_1$ and $M_2$ are the atomic weights of the incident ion and the target material, respectively. $C_m$ and m are constants that depend on the energy range. The constant $C_m$ is given by:

$$C_m = \frac{\pi}{2} \lambda_m a_{TF}^2 \left( \frac{2 Z_1 Z_2 e^2}{a_{TF}} \right)^{2m} \left( \frac{M_1}{M_2} \right)^m,$$

where $Z_1$ and $Z_2$ are the atomic numbers of the incident ion and the medium, respectively, e is the electronic charge, $a_{TF}$ is the Thomas-Fermi screening radius: $a_{TF} = \frac{0.0468}{\left(Z_1^{2/3} + Z_2^{2/3}\right)^{1/2}}$ nm. The value of the constant $\lambda_m$ depends on m, which is further determined by the value of the reduced energy $\varepsilon$. The reduced energy depends on the ion energy and is given by:

$$\varepsilon = \frac{0.03255 E(eV)}{Z_1 Z_2 \left(Z_1^{2/3} + Z_2^{2/3}\right)^{1/2}} \frac{M_2}{M_1 + M_2}.$$

Given the reduced energy, one can determine the value of m and $\lambda_m$ by utilizing the following correspondence: 1) m=1/3 and $\lambda_m$=1.309 for $\varepsilon \leq 0.2$, 2) m=1/2 and $\lambda_m$=0.327 for $0.08 \leq \varepsilon \leq 2$ and 3) m=1 for $\varepsilon \geq 10$ [37]. Note that the equation for the stopping distance is more accurate at different ranges of the reduced energy. Typically the range of nuclei recoiling from WIMP interactions is O(1-10) nm.

In Appendix 4, we discuss the energy loss of background particles --- alphas, betas, and gammas--- moving through the detector; these could produce spurious signals in the detectors as they could be misidentified as due to WIMPs. There we show that the range of these particles is much longer; this difference in range may be useful in distinguishing signal from background: the recoiling nuclei from WIMP interactions stop within tens of nanometers while the background particles travel much further, at least several hundreds of nanometers. Whereas in traditional detectors, the energy resolution is not there to make the distinction, in the nano-scale detectors proposed here, this difference in range is extremely useful in eliminating backgrounds.

**Limitations of existing detectors:** The goal is to obtain the track of the recoiling nucleus after it has been hit by a WIMP. Yet in existing detectors the track length is shorter than the resolution of the detectors. Nuclei with high atomic number A also have high atomic charge Z; the density is higher, d > 3 g/cc; and the energy deposition is proportional to $Z^2$ x d. However, in detectors we proposed density may be as high as d= 19.3 g/cc. Thus the range of recoiling nuclei is super-short, often below 10 nm, while existing detectors have spatial resolution of a few microns. Thus, in both typical solid state detectors as well as liquid detectors, the range is 100 times shorter than the spatial resolution. As a consequence, in prior designs of "directional detectors", the density of the detectors must be brought low enough to increase the recoil range. For example, it is proposed to use Xe gas pumped to 0.1 Atmosphere [36,38]. Such a huge volume of gas ($10^4$ m$^3$) must be placed underground and shielded against radioactivity. Here we propose smaller and much less expensive alternatives, taking advantage of nanometer tracking.

**DM Detector Designs:** There have been about 50 different designs of DM detectors, of which about 10 were actually implemented. The first were ultra-pure semi-conducting Ge detectors [8,10], which gave some limits on DM candidates. Actually, this type of detector, developed by the team of F.T. Avignone and R. Brodzinsky, provided the best limits on the existence of DM candidates for the initial 10 years. It even provided first "traces" of annual modulation effect (AME) but at the 2 sigma level. Currently, DM detectors can be divided into the following main classes:

(I) ionization based detectors
(II) scintillation based detectors;
(III) cryogenic bolometers;
(IV) superheated detectors, including the nano-explosive detectors; and
(V) biological detectors.

Types (I), (II) and (III) have been analyzed in a large number of papers. In the recent mini-review [25] of some of us, we analyzed the limitations of these detectors for $M_{DM} < 10$ GeV/c$^2$ and concluded that the main constraint is kinematics. For CDMS and other germanium-based experiments with $E_{th} \approx 5$ keV, the smallest WIMP velocity capable of producing recoils with $E > E_{th}$ is $V_{min,th} = O(500$ km/sec$)$ for these light WIMPs, while for XENON100, LUX, and other xenon-based experiments with $E_{th} \approx 3$-$5$ keV, $V_{min,th} = O(450$-$550$ km/sec$)$. This is well above the typical WIMP speeds in the galactic frame. Thus CDMS-Ge and Xe based experimental constraints are very halo model dependent for low mass DM candidates.[1]

Because of the similar physics of our bio-calorimeters, we provide some comments about already implemented superheated particle detectors. Superheated superconducting colloid detectors were applied to the DM detection [26]. This is a class of detectors, which are both cryogenic bolometers and superheated detectors. It has many promising features, such as the "one and only one grain" principle. When operating in $^3$He or in dilution refrigerators, they can have $E_{th}$ as low as 0.1 keV. For many technical reasons, the most promising implementation will be Be- superconducting colloid detectors.

There are many possible implementations of thermodynamically superheated detectors [27]. They all can be characterized by $E_{th}$ and $E_{amplification}$, where $E_{th}$ is the energy threshold of the detector. The classical examples are bubble and cloud chambers. Most of these implementations use a superheated change of state from liquid to gas. Technically, the easiest to use is Freon (chlorofluorocarbons), which was used in PICASSO, SIMPLE and COUPP. Each chlorofluorocarbon molecule contains carbon and fluorine, and it serves as a reasonable detector for $M_{DM} = O(10$ GeV/c$^2)$. Chlorofluorocarbons do not look promising for $M_{wimp} = O(5$ GeV/c$^2)$ because of relatively high $E_{th} = O(5$ keV$)$. PICO (the bubble chamber program formerly known as COUPP is currently running with a 3 keV threshold and should have better sensitivity to lighter WIMPs. The advance that allowed this was to switch target fluids to $C_3F_8$ from $CF_3I$ [29]. The PICO team claims the new fluid is less sensitive to gamma rays at these low thresholds. Due to the presence of $^{19}$F with its well-defined spin, they may be very useful when one targets spin-dependent interactions [28]. They are based on the change of state from liquid to gas, and their spatial resolution is comparable to the bubble size of about 1 mm.

Enzymatic reaction detectors (ER-detectors) can be classified as superheated detectors. Energy deposited by recoiling nuclei leads to a change of state, so that we have a well defined $E_{th}$. Then the enzymatic reaction is triggered which amplifies the signal. Such detectors permit the implementation of new methods necessary to reject backgrounds. Furthermore, ER-detectors can be relatively large, for example, up to 1,000 kg of low A material, but the active component, *i.e,* an enzyme, is at the level of 0.1-0.5%. Therefore, less than 1 kg of enzyme will be needed. They will be useful for the detection of low mass DM particles. The enzymatic amplification detectors are based on frozen $H_2O_2$ /$D_2O_2$ mixture doped with enzyme catalase and can be seeded with passivated nano-spheres of Li, B or Be compounds. Thus, they may be very low cost detectors.

Other example of biological detectors is Au-ssDNA we proposed elsewhere and described in the following [32]. It has many advantages, but the active and expensive material, ssDNA, accounts for a few percent of the weight. Thus, enzymatic reaction detectors may be much cheaper than ssDNA based detectors.

---

[1] Lower threshold CDMS analyses exist (down to 2 keV), but such analyses have a much larger background contamination than traditional analyses [18]. SuperCDMS [18], the next iteration of the CDMS experiment, might be able to lower the threshold without background contamination.

*Low Mass WIMP detectors*: Next generation low mass DM detectors need :
(1) to be based on low A elements;
(2) to have excellent $E_{th}$ for recoiling nuclei, *i.e.*, $E_{th} << 1$ keV;
(3) to provide a few nanometer granularity or a few nano-meter spatial resolution to characterize the track of the recoiling nucleus after it has been hit by a DM-particle.

Optimal detectors for low mass $M_{DM}$ are not currently available. We propose a new class of detectors which may achieve these performances.

Nuclei with a high atomic number A also have high atomic charge Z and high density, say >3 g/cc. The energy deposition is close to $Z^2 \rho$, if the recoiling nuclei is totally stripped. Thus the range of recoiling nuclei may be super-short, often below 10 nm for high A, but would be O(50 nm) for low A. The best existing detectors have spatial resolutions of a few µm. Thus, in existing detectors, the recoiling nuclei track length is much shorter than the spatial resolution of the detector. For example, in typical solid-state detectors, the range is 100-fold shorter than the spatial resolution. Thus existing detectors cannot be used as vertex detectors.

For granular detectors, a DM-particle leaves most of the energy at the grain with which the DM particle interacted; this energy leads to a change of phonon spectrum, *i.e.* the grain is heated. Thus, the energy deposition in the vertex grain may be much larger, up to a factor of ten, than in the next grain crossed by recoiling nuclei. This vertex voxel advantage is lost in liquid or gas detectors; the spatial resolution is defined not by granularity but by the range of recoiling nuclei in a given medium.

In a recent paper [25], some of us characterized detector properties to satisfy certain SIGNATURES. We stressed the need for detectors with $E_{th} < 1$ keV to enable the detection of DM candidates with velocity smaller than $V_{peak}$. The energy deposited, $E_{deposited}$, and dE/dx are dependent on material Z and density and on assumed $M_{DM}$. For $M_{DM}$ = 0.5, 1, 5, 10 and 15 GeV/c$^2$ we calculated kinetic factors; estimated the range of recoiling nucleus and compared it to the range of single charged particles and alphas. This permits an evaluation of the ability to reject background. When compared with the spatial resolution of the detector, it offers the confirmation of the direction of the incoming DM-particle. There are two possibilities based on either a clear distinction between head/tail of the track of recoiling nucleus, or that the range of recoiling nucleus is at least five-times greater than the spatial resolution of the detector.

To maximize the energy transferred, we need deuterium or helium based detectors if $M_{DM}$ =O(3 GeV/c$^2$ ) and Li, Be, or B based detectors if $M_{DM}$ = O(5 GeV/c$^2$). The mixed material may be used to maximize both the energy transfer (low A) and stopping power for recoiling nuclei (high Z, high density). Interaction on (C, N, O) are marginally acceptable for $M_{DM}$ = O(5 GeV/c$^2$) if $E_{th} < 1$ keV. Importantly, even if cross-sections of the interaction with DM candidates fall fast because low A materials need to be used; it is not the signal, but signal to background (S/B), which is limiting.

The best way to reject background is to measure $E_{deposited}$ in a vertex. Until recently these vertex measurements were not possible. The typical range of recoiling nuclei is much smaller than the spatial resolution of existing solid-state or liquids-based detectors. For example, an expected energy transferred to oxygen by DM ($M_W$ = 5 GeV/c$^2$, V= 220 km/sec) is about 0.5 keV. Then, the range of an oxygen ion in water is about 10 nm. Thus, almost all energy is deposited in the vertex, *i.e.* within R= 10 nm. This should be compared with energy deposition by a single charge relativistic particle wherein only 1% of energy is deposited in the vertex. Similarly, for a typical alpha particle, $E_{alpha}$ = O(2 MeV), but the range is about 50 µm. This means that the energy deposition is about 1.0, 0.01 and 0.001 keV for recoiling nuclei, alpha particles and minimum ionization particles, respectively. Thus, to reject background(s), bolometric detectors of DM with nano-size granulation must be based on $E_{deposited}$(vertex) and not on $E_{deposited}$ (all tracks).

## 3. NANO-METER SIZED BOLOMETERS.

There are many possible implementations of thermodynamically superheated detectors [27]. Table 2 in the companion paper on nano-booms detectors shows characteristic properties of such detectors, which can be divided into four sub-groups: classical superheated detectors based on properties of certain liquids (mostly based on freon), superheated superconducting colloid, enzymatic-reaction detectors and nano-formulation explosive detectors [25].

The use of bolometry is very favorable in case of DM detection; *i.e.*, $E_{deposited}$ in the vertex is larger and better defined than dE/dx over the track. If the detector is solid-state, a large fraction of recoil energy goes to the lattice and is degraded as phonons *aka* heat. The dE/dx of recoiling nuclei may be much larger than that of the single charged particles background, *e.g.* betas and CRs

Thermodynamic conditions, *ergo* the possibility of triggering a superheated system are similar in both well understood "classical" thermodynamically superheated detectors; including bubble chambers, cloud chambers and SSC; as well as "new" detectors, such as nano-explosives and enzymatic reaction-based detectors. As a class, the thermodynamically superheated detectors are characterized by a certain $E_{threshold}$ and by $E_{amplification}$. For ER-detectors, $E_{threshold}$ depends on the enzyme processivity *vs.* temperature curve. The energy amplification depends on the enzyme rate, titration of enzyme and some properties of substrates. Optimization of these parameters is necessary to make ER-detectors suitable for the detection of low mass DM for both spin dependent and independent interactions.

To detect heat effects in very small objects, one needs optimal amplification, which, for case of nano-explosives, can be as high as a factor of ten thousand. In ER-detectors, the amplification is tempered, *i.e.* O(100), which makes possible the implementation of detectors in which the signal creation is in a vertex, say a 100 nm x 100 nm x 100 nm, voxel. ER-detectors can work in water or heavy water ($D_2O$); *i.e.* DM particles interactions are with H, D and O. The ER-detectors can be seeded with nano-grains of (Li, Be, B) which may be used to optimize the range of targets. Our design of ER-detectors is conservative. Commercially available enzymes have been screened and reaction conditions are only slightly modified to design ER- detectors. Full implementation will require optimization and scale-up. Some improvements can be potentially accomplished by using recombinant enzymes.

**Nano-explosive detectors:** Before introducing ER-detectors, we need to describe advantages and challenges of nano-explosive detectors [34]. This concept has four main components: ( a) Nanotechnology permits the production of grains with R=O(5 nm), *ergo* very small specific heat; ( b) DM scatter on nuclei ($E_{deposited}$ < 1 keV)) but the signal can be amplified to about 2 MeV/grain by the release of energy stored in nano-grains; (c) Explosion initiated in a single grain propagates until terminated by an appropriately designed "heat barrier", and (d) Explosion creates a large amount of gas, which generates strong sonic effects ("nano-booms").

There are two important challenges: understanding of the ignition, and the ability to terminate the explosion at some pre-designed threshold, say $R_{explosion}$ = 100- 500 nm. Ignition is initiated by the thermal process; *i.e.* there is local heating of a material at $R_{grain}$ < 10 nm associated with the WIMP recoil event. Thus, nano-explosive detectors are nano-bolometers. The majority of explosives have been researched and formulated as µm size crystals; only in the last decade has handling of nano-explosives become popular. Explosives that are highly unstable might be stabilized if they are synthesized as small insulated nano-grains. The life-time of super-sensitive explosives is expected to scale inversely with the volume of explosive grains even if sensitivity is proportional to the grain volume.

For solid-state detectors and $M_W$ = O(5 $GeV/c^2$), the deposition of energy to the lattice dominates. Ballistic phonons are created which decay into thermal phonons (heat). Due to phonon reflection from the grain surface, the temporal scale of heating (nsec) is much faster than that heat escape (µsec). This leads to an average temperature increase of
(3.2)     $dT = E_{deposed}/(C_v(T)xV)$,
wherein
(3.3)   $C_V(T) = a(T/T_{Debye})^3 + bT$,
The first factor accounts for lattice-specific heat and second for electronic specific heat. In insulating crystals the specific heat of the lattice dominates, and in metals the specific heat of electrons dominates. We note, that this formula is adequate in the case of ER when the substrate is ice, but needs to be reconsidered when it is water.

## 4. *DNA BASED DETECTORS*

In a previous paper, some of us described the concept of a DNA-based Dark Matter (DM) detector [32]. We described a gold (Au) ssDNA detector compatible with a directional search for DM. Detectors made of ssDNA may provide nanometer resolution tracking with energy threshold below 0.5 keV. Such an Au-based ssDNA detector can operate at room temperature. Since the atomic mass of gold is A=197, gold foils would be excellent targets for high mass WIMPs. The implementation consists of a large number of thin foils of gold , from which strings of DNA hang down. The sequence of the ssDNA is known. When a WIMP scatters elastically off of a Au nucleus, the heavy ion traverses through a few hundreds of ssDNA strings before either coming to a stop or impacting the mylar film on the other side of the ssDNA layer (where it will be stopped). By interacting with ssDNA, the nucleus breaks the ssDNA strands. The fragmented strands are recovered and periodically removed. The ssDNA can be amplified by a number of techniques, such as Polymerase Chain Reaction (PCR). Then the collection of amplified ssDNA fragments becomes what biologists call a DNA ladder. It can be sequenced with a single base accuracy, corresponding to a precision of 0.7 nm for straightened DNA strands. Thus the path of the recoiling nucleus can be tracked with nanometer accuracy. A few slightly different configurations can be implemented. The important new development is the idea of using DNA in lieu of more conventional detector materials to provide a thousand-fold better tracking resolution so that the directionality of the WIMPs can potentially be determined. One of the advantages is that the detector is highly modular, and the material from which the foils are made can be easily modified.

We considered the compatibility of an Au-based ssDNA detector detectors with direction sensitive DME; *i.e.* we needed to know the count rate as a function of time, direction and energy of the recoiling ions. To measure the spherical angle with 10 degree precision, we assumed five broken ssDNA strands. To measure the energy (dE/E = O(10%)) we need 100 broken ssDNAs strands. We recently realized that, to detect the WIMPs streaming in from the direction of the constellation Cygnus, we do not need the energy spectrum [33]. We need to mechanically support the gold foil, with some material that does not interact much with WIMPs. We can do this using a graphene lattice consisting of nm wires in a quadratic array. Most of the lattice is empty space which does not interact with DM; yet the graphene has the strength to stabilize the foil, We then have a forest of DNA hanging off the graphene foil, which is distanced about 0.5 mm from a foil.

The current price of 99.9% pure Au is about \$49/g. For nucleic acids, the current cost is between \$500/g to \$2,000/g. One of us (G. Church) is developing methods, which by 2015 should reduce the price of DNA to about \$100/g. Thus, a ssDNA detector compatible with directional detection of WIMPs permits considerable reduction of material costs to below \$10,000,000/detector.

While gold has been routinely used as the proposed target material in these ssDNA detectors, due to it being an ideal target for a wide range of WIMP masses (100 GeV or larger), a variety of other target materials may be used instead. For lighter WIMPs (~ 50 GeV or smaller), a lighter nucleus would be preferred over gold as recoils of the latter would have very low energies. Possible targets include beryllium, fluorine (in Teflon), or aluminum, where the choice of target can be tailored to maximize sensitivity to particular WIMP mass ranges or coupling types (e.g. spin-dependent interactions).

Several challenges to the ssDNA design need to be resolved: the necessity of a large amount of DNA, the regular spacing of the DNA strands, the ability to use stretched ssDNA (see Appendix) as opposed to coiled ssDNA. and the ability to diminish the strand-to-strand interaction. These challenges are all under investigation and none appear to be an insurmountable obstacle.

There are many advantages to this new technology of using DNA:
1) Nanometer spatial resolution enables a directional detector with detector mass of 10 to 100 kg;
2) Detector operates at room temperature and $E_{th} < 0.5$ keV;
3) One may choose a number of low-A elements to enable the study of $M_{WIMP} < 10$ GeV/c$^2$;
4) May contain $^9$Be, $^{13}$C and/or $^{19}$F to maximize spin-dependent interaction rate;

5) Detector signal may be amplified by a factor of $10^9$ by using DNA amplification;
6) Excellent background rejection can be obtained by using dE/dx in the vertex and $> 10^{12}$ granularity.

We performed simulations, which confirmed the ability to detect the DS-DME using Au-ssDNA. Our simulations suggest that Au-ssDNA can detect infall towards the direction of Cygnus modulated by the diurnal rotation of Earth.

*Properties of the ssDNA detector.* We note that the reason we use ssDNA rather than double-stranded DNA (dsDNA) in the detector is that dsDNA is too hard to break . The mechanism of the interactions of single charged particles and/or alpha particles with dsDNA has previously been studied experimentally. The main interactions are with electrons, and the probability of breaking dsDNA is small, typically below $10^{-6}$ per interaction. Some of us have begun to experimentally perform similar experimental studies of ssDNA breakage by a variety of incoming ions to better understand the cross sections for cutting ssDNA.

When a WIMP scatters with a nucleus in the detector, it propels the nucleus forward into the hanging strands of DNA. For strands of 1000nm length placed 10nm apart, the ssDNA layer is optically thick to most recoiling gold nuclei. That is, recoiling nuclei are expected to collide with and sever O(1-10) DNA strands as they pass through the DNA layer. For recoiling nuclei exiting the foil nearly perpendicular to the plane of the foil, the DNA layer becomes optically thin: by traveling nearly parallel to the DNA strands, the nuclei are unlikely to hit one of those strands (and extremely unlikely to hit more than one strand). However, for the given strand length and spacing, the DNA layer becomes optically thin only for recoils within a few degrees of normal to the foil plane. As the nucleus can recoil over a range of directions given an incoming WIMP direction (by up to 90° relative to the WIMP), only a small fraction of nuclear recoils will be in an optically thin direction. Thus, most nuclei entering the DNA will sever multiple strands, allowing for the trajectory to be reconstructed.

*The directional capabilities of the ssDNA detector.* The 3D track reconstruction capabilities of the ssDNA detector will allow the distribution of the incoming WIMP directions to be determined. A unique signature of WIMPs is that the main direction of the WIMPs and, thus, the recoiling nuclei in the lab frame varies throughout the day as the orientation of the detector changes relative to the galaxy (due to the Earth's rotation), leading to the diurnal modulation in the signal. This signature can be detected even without the 3D track reconstruction as the overall event rate in the ssDNA modulates with the orientation of the detector. In the simplest case, suppose the detector is oriented such that the plane of the foil is perpendicular to the incoming WIMP direction, with the ssDNA layer behind the foil (i.e. opposite side of the foil as the incoming WIMPs). While the nuclei in the foil can recoil in a variety of directions, it will always be within 90° in the WIMP direction. This means that all recoils will be towards the ssDNA layer behind the foil and thus will produce a signal in this detector. On the other hand, if the detector orientation is rotated 180° so that the DNA layer is in *front* of the foil, all nuclear recoils in the foil will be *away* from the DNA region and no signal will be produced. These two cases can be easily distinguished due to simply the presence or lack thereof of severed ssDNA strands, even without using severed strands to reconstruct the nuclear recoil trajectories.

In a real detector, fixed on Earth, the orientation of the detector will not naturally vary from one of these two extreme cases to the other due to (1) the location of the WIMP wind relative to the rotation axis of the Earth and (2) the effect of latitude on the rotation of a fixed detector. However, while the signal rate will not vary between these two extremes, it will undergo a significant change throughout the day. If the detector were not fixed directly to the Earth, but placed in some orientable housing, the detector could indeed be oriented to maximize or minimize the signal. The room temperature operation of these ssDNA detectors, with a corresponding reduction in required infrastructure, may allow for just such a setup.

While not necessary for observing the diurnal modulation effect, the full 3D track reconstruction will reduce the number of WIMP interactions necessary to actually detect this effect. It will also allow for a characterization of the WIMP velocity distribution in the halo, providing insight into the structure and formation of the Milky Way.

*The background rejection of the ssDNA detector.* The ssDNA detector will distinguish WIMP-induced nuclear recoil from backgrounds such as electrons, alpha-particles and cosmic rays. We may differentiate signal from background because the recoiling nuclei from WIMP interactions stop within tens of nanometers inside the foil or inside the mylar on the other side of the DNA layer, while the background particles travel much further. Thus the recoiling nuclei will be contained within one DNA layer, whereas the backgrounds will traverse several in sequence.

Alpha particles (Z=2, A=4) will have similar type of interactions as recoiling nuclei. However, their energy is in a range of a few MeV rather than in keV. Thus, they will traverse about 50 planes and break hundreds of ssDNA. They can be almost completely rejected by the physical granularity of the proposed detector. This suppression can be as good as a factor of one million (Monte Carlo simulations are being performed). Furthermore, alpha particles are emitted over 360 degrees whereas the recoiling nuclei are in the same general direction as the WIMP. Thus, the signal caused by alpha particles is not modulated by the Earth's rotation.

The spectrum of beta particles will be dominated by $^{40}$K with $E_{max}$ = 1.32 MeV. The range of such electrons is about 100 planes. Furthermore, they will interact with electrons of ssDNA , and this leads to a smaller, (say << 1%) probability of breakage. In addition, they are emitted over 360 degrees and should not correlate with the Cygnus direction, using AME or DME.

There is a very efficient mechanism for the rejection of Cosmic rays, including muons. The main components are protons and muons, which interact mainly with electrons, wherein the probability of breaking ssDNA is very low. There are also rare interactions with hydrogen, which may lead to a break-down of ssDNA. However, in this case, the cosmic rays will have enough energy to traverse all planes of the detector. They will be rejected due to the extraordinary granularity of our detector.

An important source of background will be the interactions of fast neutrons with the foil. This topic has been discussed in the recent paper of [25]. We stress that fast neutrons from U/Th are emitted over 360 degrees, whereas the recoiling nuclei are scattered forward and generally conserve the WIMP direction. Thus, the signal due to fast neutrons should not be modulated by Earth's rotation.

## 5: APPLICATIONS OF ENZYMATIC REACTION DETECTORS FOR LOW MASS DM-CANDIDATES

In last thirty years there has been rapid progress in nanotechnology; grains with R=O(5 nm) can now be efficiently produced and passivated by a very thin layer, say O(1 nm), of another material. Thus, even materials which interact with water such as Li, B can be placed inside of {$H_2O_2$ + $H_2O$} ice. We will discuss, in the following section, a Li grain with $R_{Li}$ = O(5 nm) coated by thin layer, say 0.5 nm of Au as an example. DM particles will interact with the Li, the recoiling nuclei are stopped in Au and heat escapes to trigger an enzymatic reaction in close vicinity to the grain. Thus, the Au-coated Li grain is a transducer, and the ER is a sensor/amplifier combination. We note that grains are in the solid state and do not change state; *i.e.* only $C_v$ is of importance. In contrast, for ice, the energy budget is dominated by the thermodynamics of change of state.

For most materials, $C_v$ = $10^{-5}$ keV/(nm$^3$·°C) at room temperature. In addition, because a large fraction of recoil goes directly to the lattice, an energy deposition of 1 keV/grain is possible. Actually, when using "mixed" material consisting of both low A nuclei for DM targets and high Z (for maximizing dE/dx), we expect that O(1 keV) is deposited in R=O(5 nm) grain. Thus, in the grain of R= 5 nm$^3$ => $V_{grain}$ = 525 nm$^3$, the temperature increase may be as large as 100 °C.

The advantage of using nano-beads to seed the ER-mixture ice is that, within sub-microseconds, the energy deposited in a nano-grain is released producing a layer of liquid around a grain. The enzymatic reaction may then increase the thickness of such layer to about 0.5 µm. The enzyme selected, catalase, leads to the release of $O_2$ gas in the reaction; *hence,* a spherical wave of hot gas is produced, and this leads to a sonic boom. Furthermore, as described below, isotopically enriched oxygen can be used, allowing the detection of and $^{17}$O and $^{18}$O may be detected by mass spectroscopy. By looking at three different mass peaks, there is there is almost perfect recognition of any possible background.

**Methods of read-out.** When using catalase, most of the energy is carried out as hot gas, *i.e.,* $O_2$ molecules. This will trigger melting in the spherical layer close to the vertex nano-grain. Let's assume that we package the R=5 nm grains into a $R_{ball}$ = 100 nm. Furthermore, let us assume that the volume of transducer material, such as Li, is about 25% of the total volume. There are $N_{grains}$ = 0.25 x (100/5)$^3$ = 2,000 grains in a ball. The total energy produced by an explosion of single 100 nm ball is about 1 MeV x 2,000 = 2 GeV. This will be about 16 GeV if balls of R=200 nm are used and 250 GeV if $R_{ball}$ = 500 nm. The energy of 1 GeV/c$^2$ should be detectable by a sufficiently sensitive sonic system. Because when temperature increases above 60 °C, the enzyme denaturates, the melting is self-limiting and the maximum volume of melted ball is roughly proportional to titration of enzyme.

Let's assume that the substrate is {(10%) $H_2O_2$ + 90% $H_2O$}. In the same geometric situation, the total mass of $H_2O_2$ is 0.1·$V_{ball}$·ρ(g/cc) = 5.22 x 10$^{-16}$ g. With 50% enrichment with $^{18}$O there are about 3.2 x 10$^8$ atoms of $^{18}$O. This amount can be very easily detected by any modern mass spectrometer. *The dE/dx for single ionizing particles is much smaller.* Hence, any event induced by recoiling nuclei will be seen, whereas single-charged particles are very inefficient in initiating the enzymatic reaction.

There is a third method for detection of DM-induced creation of a zone of sustainable enzymatic reaction. The "$H_2O_2$ ice" say 500 nm thick may be placed between two graphene planes. The conductance of ice is very low. However, a water channel, created by enzymatic reaction initiated by recoiling nuclei, has a much larger conductance. Thus, there is a possibility of seeing the "short-cut" between two graphene planes. Similar techniques were developed as a bio-sensor for the presence of large macromolecules, and excellent S/B's (> 100) have been observed. This method will be described in forthcoming paper.

**Backgrounds.** ER-detectors can be triggered by recoiling particles, particularly if the granulation is on the R = O(10 nm) scale. Single charged particles are not expected to initiate the ER thermally. For relativistic single charged particles (CR, muons, fast betas), the typical dE/dx = 2 MeV x (g/cm$^2$). For ρ =O(1 g/cc) this is dE/dx = (10$^6$/10$^7$) = 0.1 eV/nm. For single charged betas at the end of their trajectory dE/dx = O(0.5 eV) and for alpha particles dE/dx = O(2 eV/nm). The probability of increasing temperature in R=5 nm by 1 °C is negligible [ P = exp(-0.5 keV/5 eV) = exp(-100) ]. Hence, the only background in such detector is expected to be due to fast neutron scattering on nuclei. Also in this case, the majority of $E_{deposed}$ is to the lattice and is immediately transferred into heat.

We should consider background due to temperature fluctuations. Temperature is a mesoscopic notion and starts to lose its meaning in aggregations below 1,000 molecules. This is why, we do not propose the use of sub-nanometer grains even if they may be produced. We believe that grains of R = 10 nm with about million molecules per grain are close to optimum. In this case, the temperature fluctuations are negligible. The external temperature stability may be as good as 0.05 °C. Thus, with assumed heating by DM interaction of about 5 °C, the two processes have very different probabilities, $P_{T-fluctuations}$ = exp(-5/0.05) = exp(-100), *i.e.* once more the expected background is negligible.

In a companion paper in this collection, we documented that nano-explosives can be triggered by energy deposition due to the interaction of DM-candidates with highly granulated high energy content material. We also calculated that a small energy deposited, O(0.1 - 0.5 keV) may lead to the release of about 2 MeV of energy stored in R=5 nm explosive grain. This may lead to the ignition of a certain number of close grains, *i.e.* to initiation of an avalanche. However, we need 10 kg detectors for $M_{DM}$ < 10 GeV/c$^2$. Thus, the main question is not if the run-away process starts, but how to terminate the run-away process at a pre-selected distance. This can be engineered by having spacer material damping the avalanche amplification and by using lower amplification to slow down the explosion. Yet, it would be desirable if, as in case of freon, we would have an amplified but self-limiting "run-almost-away" process. This may be achieved with the use of enzymatic processes.

## 6. DESIGN OF ENZYMATIC REACTION (ER) DETECTORS

Enzymatic processes provide a local amplifier, *i.e.*, a source of energy, which is released when recoiling nuclei (D, Li, Be, B) interact with energy storing material. Our ER detector is enabled by a local change of state induced by either energy of recoiling nuclei or energy release in an enzymatic reaction, which is triggered by energy deposition by recoiling nuclei. Based on calculations of dE/dx, we argue that a change of state is not possible when the radiation is electrons or X-ray, but it may be realized when heavy ions are present.

The advantages of enzymatic processes as compared to nano-explosions include:
6.1) they can be triggered by a relatively small increase of temperature;
6.2) the amount of energy released is a few times smaller than in nano-explosives;
6.3) the time scale is longer [μsec];
6.4) the enzymatic processes are self-limiting when temperature rises above some $T_{critical}$, where the $T_{critical}$ is defined by $T_{denaturation}$ of an enzyme.

Concerning (6.4) there are two competing processes. Since, enzymes are proteins acting as catalysts, temperature affects both on their structure and catalytic activity *via* denaturation process. But for any catalyst the higher the temperature the faster the catalytic activity. There are hundreds of well-understood enzymes, and for many of them low-, mid-, and high-temperature variants exists. Thus, we have to select a proper type of enzyme. In the following, we will focus on the following classes of enzymes:

6.5) liquid 1 + enzyme => liquid 2 + gas + energy
6.6) solid 1 + enzyme => liquid 2 + liquid 3 + energy

The examples of (6.5) are
(6.7) $2 H_2O_2$ + {catalase} => $2 H_2O + O_2$ + energy
and
(6.8) $2 HOCO-CH_2CH_2CH(NH_2)CO_2H$ + {enzyme} => $HOCO-CH_2CH_2CH_2NH_2 + CO_2$ + energy
wherein the enzyme is glutamate decarboxylase.

Typically, the enzymatic reactions will be triggered by a temperature increase, which leads to the change of state of substrate (melting). We provided some information about acid-producing reactions because it is possible to trigger the ER by a pH change, *i.e.,* the use of two enzymatic processes - one to trigger the reaction and other for the amplification. This possibility will be discussed elsewhere.

Most of these reactions take place in acquase solution . If the water is frozen, there is no reaction, but local energy deposition, *e.g.*, due to heavy ions, may melt $H_2O$ and then an enzymatic reaction is triggered. Because of the energy supply from the reaction, it is sustained but self-limiting because of the limited supply of components and because of denaturation of the enzyme when the temperature becomes too high. This issue is important- for the concentration of substrate too low (*e.g.* below $K_m$) the reaction rate is very low and the nano-ball of water freeze due to contact with large mass of '$H_2O_2$ ice".

Generally, the ER triggered by local energy deposition can be divided into: transient; locally self-sustaining; and globally propagating. Like in any superheated detector, these phases are induced because the energy production is proportional to bubble volume and heat escape through the surface of the bubble. However, in case of ER, the additional effects of substrate depletion and enzyme denaturation have to be taken into consideration. For ER-detectors of DM candidates, the locally self-sustaining mode seems best. Note that total energy produced (amplification) is well correlated with the energy deposited in the vertex of a DM-candidate interaction. Thus ER-detectors can be operated both in $E_{th}$ and energy resolution modes.

We discuss the simplest physical process in which local heating leads to the melting of ice, which then initiates the enzymatic reaction. For a moment we will assume that heat escape is negligible; more realistic situations of heat escape make our argument even stronger.

## 7. IMPLEMENTATION USING CATALASE

Catalase is an extremely efficient enzyme. The substrate is $H_2O_2$ in water, and its concentration can be from 1 to about 50 %. We will denote such substrate as {$H_2O, H_2O_2$} or if necessary as {(1-x(%)) $H_2O$; x(%) $H_2O_2$}. In the substrate, hydrogen can be fully or partially replaced by deuterium. Commercial sources of heavy water exist wherein the supply of deuterated hydrogen peroxide can be produced. Thus, our favorite substrate will be {$D_2O$, $H_2O_2$}. As described above, the oxygen in $H_2O_2$ may be enriched with $^{18}O$ which in natural oxygen is only 0.2%. In this case is the use of {$D_2O$, $H_2O_2$} is especially favorable because natural oxygen has $M(^{16}O) = 16$ D, $M(^{18}O) = 18$ and $M(D_2O[vapor]) = 20$ D. At low masses, a difference in mass by two units is very large indeed.

Assume that Li, LiF, B or Be nano-beads are coated with a thin layer of heavy metal -such as Au and dispersed inside frozen {$H_2O_2$, $H_2O_2$}. For example, when using Li and Au, nano-beads can be produced with $\rho = 1$ g/cc. They can be coated with linkers to which is attached the enzyme catalase. There are two processes. The first is fast but provides only about 10% - 20% of the recoil energy, *i.e.*, the escape of the recoiling nucleon. In the second process, slow, but larger DM interacts with nuclei, and this leads to the excitation of the lattice and the production of ballistic phonons, which are thermalized, and then heat escapes outside.

Both of these processes transfer the energy to the local medium and then melt a few nm diameter shell of {$H_2O_2$, $H_2O$} ice. Then catalase molecules start to work. Their catalytic turnover is about $10^8$ operations per second, and to each 5 nm low atomic number bead we can attach about 10 molecules of enzyme. This density of enzyme is sufficient to sustain the melting process. A bubble of $O_2$ is generated, and this generates heat, denatures the enzymes, and prevents inflow of liquid.

We observed that in 20% $H_2O_2$, this leads to the creation of sub-millimeter bubbles. Then, as in any process of cavitation, a sonic boom is created that can be detected a long distance from the detector. This process generates about 10-fold more energy than superheating of freon; *i.e.*, both the optical and acoustic read-outs can be used. This mades plausible an elegant mass-spectroscopy based read-out scheme.

*Triggering Enzymatic Reaction*: We will calculate the effect of energy deposition by heavy ions on $D_2$ or O in a {$D_2O + D_2O_2$, catalase} based ER-detector. We will also calculate the energy deposed in a {$H_2O$; $H_2O_2$, catalase} seeded with Li/Au nano-grains. In Table 3, we provide data for some types of catalase-based reaction using LiF, B, Be and their compounds. The natural targets for DM interactions are H, D, Li, B, Be, O, and F. Table 3 provide the estimate of energy, energy $E_{recoil} = \mu^2 v^2/(2M)$, deposed by $M_W = 5$ GeV/$c^2$ with $V = V_{peak} = 300$ km/sec in these targets. Note that the energy $E_{rec}$ is approximately the typical energy given to a nucleus after interacting with a galactic WIMP. The detectable events are defined as events in which the energy transfer is > 0.5 keV.
The Table 3 shows calculated dE/dx for recoiling nuclei normalized to a recoiling Hydrogen moving through water (dE/dx(H)=0.0135 keV/nm). The stopping powers for the different nuclei moving through their respective medium are calculated utilizing SRIM [39].

**Table 1: Comparison of stopping power for different materials.**

| Detector | Be-ssDNA | ER1 | ER2 |
|---|---|---|---|
| Target | Be | $H_2O$ | $D_2O$ |
| Ion | Be | O | O |
| $\rho$(g/cc) | 1.9 | 1.0 | 1.1 |
| A | 9 | 16 | 16 |
| dE/dx | 9.6 | 9.2 | 9.3 |

Here ER1 is a detector based on 10% $H_2O_2$ in 90% $H_2O$, and ER2 is 10% $H_2O_2$ in 90% $D_2O$.

We also considered ER3 which is a {10% $H_2O_2$ +90% $H_2O$} seeded with 10% by volume of passivated Li grains, as wqell as ER4 which is a {10% $H_2O_2$ +90% $H_2O$} seeded with 25% by volume of passivated Be nano-grains, and the ER5 which is a {10% $H_2O_2$ +90% $H_2O$} seeded with 25% by volume of passivated B grains. Finally, we calculated also ER6 which is a {10% $H_2O_2$ +90% $H_2O$} seeded with 25% by volume of passivated LiF grains. We note, that LIF based ER-detector may be 1$^{st}$ to be deployed, because specific heat of LiF is lower than of Li, Be or B.
The Lindhard formula for calculation of dE/dx in different cases of importance for DM detection has been discussed in Section 3. Their applications leads to the results presented in Table 4.

**Table 2: Energy deposed to different targets in ER-detector**

| Target | H | D | Li | O | F |
|---|---|---|---|---|---|
| A | 1 | 2 | 7 | 16 | 19 |
| $A^2$ | 1 | 4 | 49 | 256 | 361 |
| $E_{recoil}$ (keV) | 0.33 | 0.50 | 0.61 | 0.47 | 0.43 |

There is a much larger probability of DM interacting with O or F, but for $M_{DM} = O(5 \text{ GeV}/c^2)$, this leads to a very small energy transfer. In the following, we assume that the $E_{th} = 0.5$ keV, thus mainly DM interactions on D and (Li, B, Be) are detected. More specifically, we have two types of events:

7.1) DM scatter on D and energy is shared by the lattice of $\{D_2O, H_2O, \text{Catalase}\}$-ice;
7.2) DM scatter inside a nano-grain (Li, B, Be, F) and energy is deposed in a thin layer of Au coat.

In the first case, we have a prompt signal, because the heat is quickly, say $< O(100$ nsec$)$, thermalized. In the second case, energy is rapidly thermalized in the nano-grain but then slowly escapes, leading to the melting of the $\{D_2O, H_2O_2\}$ nano-ice ball and the initiation of reaction. In both cases, the energy deposited is $O(0.5$ keV$)$ and the temperature increases to about 20 °C in a nano-grain of 10 nm$^3$. Catalase with a reasonable turnover at 20 °C should be used. .

Because the low difference between $(T_{trigger} - T_{operation})$, the contribution to the energy balance due to specific heat is smaller than the heat of the change of state for melting. For a full calculation, a local heating model has to be established, where the nucleation volume is smaller than the size of the ER-grain. At temperatures below 50 °C, the lattice specific heat of heavy water ice is small compared to the change of phase heat. Also because of the very high Debye temperature of LiF, its specific heat is negligible compared with specific heat of ice. Thus, first a nucleation zone is created, *i.e.,* a shell close to surface of LiF. This shell consists of water/hydrogen peroxide with T = $O(20$ °C$)$, wherein catalase acts on hydroxide and produces additional heat. This amount of heat is about 100-fold greater than heat due to the WIMP. Now, the total amount of heat released is about 10 MeV, which may lead to the ignition of the closest grains and cascade propagation. However, in contrast to nano-explosions, the avalanche in enzymatic reactions is self-limiting; when the temperature rises above say 50 °C, the enzyme denaturates.

Signal Amplification and avalanche self-termination: When an enzymatic reaction is triggered, about 1.5 keV/(nm$^3$), is produced by DM interaction and subsequent activioty of catalase inside a melted shell. The ignition is a localized and multi-step process; therefore we expect that the $\{D_2O, H_2O_2, \text{catalase}\}$ nano-ball with R=$O(10$ nm$)$ => V = $O(2 \times 10^4$ nm$^3)$ will be melted and starts to produce energy. At this stage, the signal amplified by the energy production is about 20 MeV per a single interaction of WIMP. However, an additional amplification is expected. Triggering of one R= 10 nm nano-grain of "ER-ice" leads to avalanche melting/triggering of all neighboring grains. Obviously, we need to terminate such process at a reasonable distance, such as 500 nm. The nano-grains of the ER-systems have to be separated by a thermal barrier, for example, a material which changes state at the selected temperature, slightly higher than $T_{trigger}$. This may be a layer of ice which is not admixed with hydrogen peroxide. Another option is to use materials which are sublimating with a sublimation threshold at the selected temperature, for example, at 50 °C. Thus, we expect that the "ER- grains" of about 10 nm will be spaced by a distance of 20 nm, center to center inside of R=500 nm balls. They will be capped by a 50 nm layer of enery absorbing material. Obviously it could also be a metal shell with high density, such as W or Pb. Preliminary calculations suggest that each ER-ball can be thermally insulated, so that there is only localized avalanche propagation.

In the ER-detector we considered, low ignition temperature is achieved by means of the enzyme that has negligible activity at 0 °C but reasonable activity at 20 °C, e.g. using catalase, obtained from organisms living in low temperature and has the temperature optimum at 15 - 20 °C. In the future, it should be possible to genetically engineer catalase, such that its activity has a very sharp temperature profile.

There are many other promising implementations of ER-detectors that are triggered by energy deposited by interactions of DM candidates on low A materials. They will provide the signal amplification from, for example,0.5 keV to about 10 GeV. Such amplification, however, may be too large and may lead to global rather than local ignition. We may need to use additional methods to temper/control the run-away character of proposed ER-detectors. Our choice of the catalase-based enzymatic reaction is because it is simple, robust and reasonably well understood. It is worth of exploring if there are other enzymatic reactions which are self-limiting; *i.e.*, the amplification is dumped at a certain temperature by using the enzyme that denatures at temperature lower than catalase.

Finally, we had considered a "mixed system", in which the detonator is a "nano-explosive". The first step is the ignition of nano-explosion(s), *i.e.,* the amplification of recoiling nuclei ($E_{deposited}$ = O(1 keV)) by release energy of explosive 4 keV/nm$^3$. Thus, we expect the amplification in a single nano-grain from $E_{deposited}$ = O(1 keV) to $E_{nano-explosive}$ (R=5 nm) = 2 MeV. Subsequently, this energy escapes into "ER-ice" and melts it, and the second step of self-terminating amplification (from 2 MeV to 10 GeV) is by means of enzymatic reaction, *e.g.*, catalase-catalyzed reduction of hydrogen peroxide. This mixed system(s) will be described elsewhere.

**Towards optimized ER-detector.** The enzymatic reaction systems use biological, *ergo* close to optimal, sub-systems. As seen with a variety of biological processes, there are a very large number of possible ER processes or their combinations. However, these processes are being used in non-traditional field of particle detection. Thus, there is a potential mismatch of biological processes and non-biological applications. Essentially, the optimization process is to remove this mismatch. The following factors need to be considered:

a)   selection of optimal enzyme/substrate combination;
b)   systems with/without granular "transducer", such as LiH or LiF;
c)   grain-based *vs.* continuous *vs.* mixed system;
d)   natural *vs.* genetically engineered recombinant enzyme;
e)   system with symmetric *vs.* asymmetric activity *vs.* temperature dependence.

*Concerning (a):* The selected enzyme should be well understood, simple, robust, low cost and with high turnover. The natural choice is catalase, which is the enzyme with highest turn-over known - a single molecule of enzyme processes up to $10^8$ molecules of $H_2O_2$ per second. The substrate, *i.e.,* hydrogen peroxide, is produced in millions of kg/year and costs about $1/liter. Because it is used as medication, it is sterile and features low levels of contaminants. Peroxid ($H_2O_2$) is produced by oxidation of water - water with levels of purity of U/Th smaller than 1 ppt has been produced. There is no reason why $H_2O_2$ with similar purity cannot be produced. Similar turnover and even lower level of impurities is expected for $D_2O_2$. This is important because of need of suppression of radioactive background.

Catalase can be found in organisms living in low, medium and high temperatures. For catalase from organisms that live at low temperatures, the optimal temperature of activity is shifted towards 10 - 15 °C. However, even more important than the optimal temperature is how steep the activity-temperature relationship is in a low temperature range. Curve shapes as close as possible to a step function are required. Catalase is just an example of high processivity enzyme. It may happen, that other enzymes have activity *vs.* temperature curves which are more appropriate for our tasks.

*Concerning (b):* The simplest system is a catalase-based, homogeneous system, *i.e.,* a bath of {x(%) $D_2O$, (1-x (%)) $H_2O_2$, catalase). We tested systems with x =3% and x=20%, but up to 70-80% of hydrogen peroxide can be used. Such a system is surprisingly cheap, even taking into account the need for a high activity catalase – we estimate the cost of the system to be O($10/liter) for a system not containing $D_2O_2$. The cost should increase to about $100/liter when $D_2O$ is used. Taking into account the reduced mass, such a system works best when $M_w$ = O(7 GeV/c$^2$), wherein the interaction with oxygen dominates. If $M_W$ = O(3 GeV/c$^2$), the interactions with deuterium dominate. The limitation is that one needs to start with {x(%) $D_2O$, (1-x (%)) $H_2O_2$, catalase) ice and adjust the amount of catalase so that after propagating to a few micrometers, the enzyme self-terminates the process by thermal denaturation. Also, there is an energy price for starting from ice but S/B should be excellent.

There are many methods in which nanometer-sized grains of ice are produced. Such a granular implementation means that we can use any amount of catalase - the process will be self-terminated both by denaturation of the enzyme and heat equilibrium like in superheated Freon, but could also make substrate balls of radius of about 1 μm. Because the process is initiated in a well defined nano-grain of substrate ice, a better definition of $E_{th}$ is achieved.

The use of either D or O leads to a double-peaked distribution of cross-section as function of $M_{DM}$. It may be useful to use the "transducer" nuclei, such as Li, Be, and B which are best matched to light WIMPs. However, these are metals with relatively high specific heat. Thus it may be better to use inorganic nano-grains, including LiH, LiF, BN, and $BeO_2$. Another possibility is to use $(CH_2)_n$ or $(CD_2)_n$ polymers. Such transducer grains, such as LiF, can be ordered in a dense array and the space between grains filled up with {$H_2O$, $H_2O_2$, catalase}. This will be by far the cheapest implementation, because low cost Li replaces the more expensive deuterium as a target for lowest mass WIMPs. In this case, catalase will be attached by appropriate linkers to "transducer" nano-grains. Note that the mixture of water and hydrogen peroxide can be replaced periodically.

*Concerning (c)*: The main concept of very large thermal engineering (VLTE) is that one can define the heat escape from a heat source. If the material is homogeneous, this leads to dynamic, spherical heat propagation, where the temperature in the vertex is highest. However, if the material is close to a change of state transition, the typical situation will be that the vertex material melts but, at some distance $R_{crit}$, the temperature is too low to change state.

This description should be modified, if the heat production by enzymatic reaction is triggered by a change of state. Actually, we demonstrated experimentally that this may lead to self-sustaining topology, where the central reaction zone is active until all energy sources are used. In the case of a catalase based reaction, this will be a few minutes, necessary to significantly deplete the amount of $H_2O_2$ in the vertex. Thus, the expected time scale is:

$t < t_1$         Thermalization of ballistic phonons created by particle interaction in vertex;
$t_1 < t < t_2$     Creation of a zone of melted substrate and stabilization of enzymatic reaction in the zone;
$t_2 < t < t_3$     Depletion of substrate ($H_2O_2$) and termination of the enzymatic reaction.

For catalase-based reaction, we expect $t_1$= O(1 nsec), $t_2$= O(1 µsec) and $t_3$ = a few minutes.

It is crucial to know how fast ballistic phonons thermalize. For example, it may be better to have ice nano-grains inside of a certain fluid so that the phonon-mismatch is established. Thus, during initial nanoseconds, ballistic phonons bounce from the surface of the substrate ice, and all energy deposed by particle leads to melting, thereby creating a larger melt zone.

The similar consideration hold when we have a substrate doped with nano-grains of transducer material, such as a compound of (Li, Be, B). All the energy is thermalized in such grains, and then a melting of substrate starts in a shell around such a transducer nano-grain. This leads to the initiation of enzymatic reactions.

*Concerning (d):* Selection of natural *vs.* genetically engineered enzyme is an important step of optimization. We may opt for a system in which energy is deposed in "ice" or in water with T slightly above 0 °C. In the first case, the "background" is negligible; catalase does not work on $H_2O_2$ ice. However, the particle has to depose more energy because the heat of melting is higher than heat necessary to increase the temperature of water by, for example, 10 °C.

The ideal enzyme, would have activity close to a step function with threshold of about 10 °C. Then, the substrate/enzyme mixture is liquid and kept at 1 °C, and the reaction is initiated when the temperature rises above the $T_{threshold}$ (reaction). Optionally, we could operate at slightly above melting temperature and use high temperature catalyse with peak of turnover at 60 °C. The heat necessary to heat water by say 50 °C is smaller than to melt ice.

In natural enzymes, the activity *vs.* temperature function is a S-shaped curve. It will be an interesting "data-base" search to find an enzyme, preferably catalase, with steepest rise of activity *vs.* temperature curve. Genetic engineering permits to modify properties of enzymes. It is not so probable that we can improve on the turnover of a super-enzyme" such as catalase. But it may be possible, to make the activity *vs.* temperature curve sharper.

*Concerning (e):* The majority of enzymes have a bell-shaped activity *vs.* temperature curve. We have already discussed advantages of as-sharp-as-possible on-set of enzyme activity. However, if we want to self-terminate the enzymatic "run-away" reaction, we need an activity curve with a reasonably sharp trailing arm. This is accomplished if the particular enzyme denatures in a well defined temperature. However, in nature, denaturation is described by S-shaped curve. The catalase from different organisms may have different width of denaturation which may be O(5 °C).

## SUMMARY

The next generation of detectors for $M_{DM}$ = O(5 GeV/c$^2$) will require: large masses, say O(100 kg) of material with low A; much better energy sensitivity, $E_{th}$ < 0.5 keV; as well as better background rejection. In the second generation, detectors with directional capabilities will be implemented. These requirements may be satisfied in the proposed "enzymatic reactions" detector", which are thermodynamically superheated detectors.

Our recent analysis of existing experimental results suggests that $M_W$ = O(5 GeV/c$^2$). For low energies there are three main reasons why existing detectors or their "extensions" are not adequate:
(1)  kinetic considerations require $E_{th}$ < 500 eV and/or use of materials with A < 10;
(2)  for coherent interaction, low A => the count-rates goes down and mass of detector is an important limitation;
(3)  in all already implemented detectors, <u>all</u> methods of background rejection become much less effective.

No existing detector satisfies all of these conditions, so that we have designed new class of detectors that involve nano-explosives and biological detectors, such as ssDNA-based detectors and ER-detectors. We have demonstrated that a simple catalase-based detector is promising, but considerable improvements can be implemented by engineering the enzyme properties and reaction process.

## ACKNOWLEDGMENTS

MC and WW acknowledge support from the Wyss Institute at Harvard. KF gratefully acknowledges support from DoE grant DE-SC0007859 and from the Michigan Center for Theoretical Physics. CS thanks Pearl Sandick and the University of Utah for support. We are grateful to Dave Gerdes, Rachel Goldman, Myungkoo Kang, Cagliyan Kurdak, Mykola Murskyj, Jordan Rowley, and Greg Tarle for useful conversations.

## REFERENCES (Dark Matter):

---

**REFERENCES (Enzymes, especially Catalase)**

[E13] "Peroxi Base - The Pereoxidase Database", iSwiss Institute of BioInformatics.

**Appendix 1: Towards practical realizations of a ssDNA detector**

**The Art of Stretching DNA:** The use of stretched strands of DNA to provide high-resolution, directional, and inexpensive particle detection could provide the next generation of powerful Dark Matter detectors. While previous single-molecule studies have successfully demonstrated the mechanical stretching of single-stranded DNA (ssDNA), achieving a homogeneously stretched layer of ssDNA over a macroscopically large area will be substantially more challenging. In this appendix, we discuss how these technical challenges could be surmounted, leading to practical experimental realizations of a stretched ssDNA detector, The successful creation of such a detector involves three basic steps which we outline below.

**A. Preparation and stabilization of ssDNA**: The formation of secondary structures in detector ssDNA will make much more difficult the accurate reconstruction of particle tracks. Most directly, secondary structures can be removed through the application of approximately 15 piconewtons (pN) of mechanical force [A1.1-A1.3]. Alternatively, previous studies have used glyoxal to chemically stabilize ssDNA and minimize the formation of secondary structure, reducing the force required to extend ssDNA to approximately 60-80% of its contour length to about 10 pN [A1.10]. Glyoxal inhibits base-pairing by reacting with hydrogen-bonding moieties on DNA bases [A1.11], although we note that glyoxal forms glyoxylic acid in the presence of oxygen, which degrades DNA. Alternatively, the use of a custom sequence without secondary structures – such as a sequence that contains only two or three different types of bases – would avoid the problem, although the use of such a sequence may be more expensive.

**B. Packaging ssDNA**: The angular resolution of the detector depends on the packing density of the ssDNA in the detector. As discussed in Section 3, recoiling Be nuclei may cross about hundred strands of ssDNA. Approximately five ssDNA severing events are needed to reconstruct a particle track. By considering the backbone of DNA as an extended chain of carbon, and balancing the ionization energy of DNA with the Coulomb energy associated with, say, an incident Be nucleus, we estimate that a metal nucleus must pass within $\delta \sim 0.1$ nm of the DNA backbone to cause a break. This estimate is of the same order of magnitude as the covalent radius of carbon. Assuming uniformly distributed random particle incidence on a square lattice of DNA strands spaced $a=10$ nm apart, we would then expect a break in $\delta/a = 1\%$ of unit cells on the lattice. Thus the required ssDNA length is approximately $N*a^2/\delta = 5\ 000$ nm (8500 bp) in order for particles incident at $\pi/4$ radians to fulfill this condition, where $N = 5$ is the desired number of severing events. As the required length varies inversely with the ssDNA area density, ~125 μm length strands are required for DNA spaced approximately 50 nm apart.

Entropic and electrostatic effects are the primary obstacles to close ssDNA packing in the detector. The mutual electrostatic force per unit length due on two unshielded charged ssDNA *(i.e.* one charge per base) strands at 10 nm spacing is approximately 0.1 pN/pm. This force is much greater than the force required to mechanically stretch it. By comparing the energetic cost of fluctuations in such a charged chain to the thermal energy in the system, we find that the lateral fluctuations of the chain are on the order of 0.5 nm. Thus, if DNA packing at sufficient density can be achieved in pure water or in neutral gas, the achievement of stretched of DNA may be possible. We would even expect that severed strands will be ejected from the layer, which is convenient for their collection. However, the presence of charges in solution will decrease this force, and we should consider the effect of drying on the charge of the ssDNA. In these two situations, external force may be required to extend the DNA.

**C. Methods of force application**: To stretch an array of ssDNA mechanically , a scalable method of parallel force application is required. Recently, a number of parallelizable single-molecule force techniques have been demonstrated, using centrifugal force [A1.4], magnetic fields [A1.5], direct displacement of a force probe array [A6], and gravity [A1.7]. In fact, multiplexed force spectroscopy experiments at the level of thousands of parallel experiments has been demonstrated by one of the authors by mounting an array of DNA-tethered beads to a rapidly rotating microscope [A1.4]. Applying force using gravitational and centrifugal force fields is convenient for our current application, due to the ease with which these methods scale to increasing numbers of tethers.

Design specifications require a minimum force of 10 pN and a strand spacing of approximately 50 nm. This force is achievable by using 50 nm Au beads (19 g/cc), and a relative centrifugal acceleration (RCF) of approximately 1,000,000 G—this is not only within current engineering capabilities, but can in fact be achieved with commercial benchtop ultracentrifuges (*e.g.* using a rotation arm with length = 40 mm and a rotation rate of approximately 150,000 RPM). A longer arm or slower rotation rate would cause the required bead size to limit the array density, but sufficiently long strands of ssDNA would compensate for these issues by increasing the probability of a break. Alternatively, we could attach many beads to each ssDNA strand, say every 100 bases, to achieve both high packing density and a sufficiently high stretching force (*e.g.* 125 gold beads of diameter 10 nm would generate the necessary stretching force at 1 million G).

An alternative approach to creating an array of stretched DNA is to generate a self-assembled monolayer of ssDNA, where the mutual electrostatic repulsion of adjacent strands is sufficient to extend the ssDNA away from the surface. Self-assembled ssDNA mono-layers of ssDNA have been successfully created by using Au-thiol chemistry with density $10^5$/cm or better [A1.9], but can exhibit substantial heterogeneity [A1.8]. While these strands are shorter than for bead-based force stretching approaches, the higher density should provide sufficient angular resolution and a high enough detection cross-section to achieve the desired sensitivity. Importantly, the collection of severed strands would be different in this case: the strands, ejected from the monolayer due to their charge, would be collected by a fluid/gas flow or on a charged surface for sequencing.

Macromolecules Vol. 4 Iss. 7(pp 2328-2333).

[A1.11] McMASTER, G. K. & CARMICHAEL, G.G. (1977). Analysis of single- and double-stranded nucleic acids on polyacrylamide and agarose gels by using glyoxal and acridine orange. Proceedings of the National Academy of Sciences, U.S.A. 74, 4835-4838.

**Appendix 2: A few remarks about the enzymatic reactions.**

At this stage of development of ER detectors we need:
1) To understand the turnover rate *vs.* Temperature (the S-curve onset and fast fall at $T > T_{crit}$);
2) Choice of enzymes from low temperature organism (example of Catalase);
3) Study of trade-off(s): starting with ice and melting (no background, but needs much energy) *vs.* (starting at say 1 °C but there is some background activity);
4) Study if the sharpness of onset of catalytic mechanism the same in all biological enzymes - are there some "miracles"?

The main challenge, *i.e.* "making sure that enzymatic reaction starts only at a certain, pre-selected temperature is similar to the requirements on "hot start" in PCR. There are at least three ways to achieve "hot start" in PCR, potentially applicable to other enzyme reactions:

(a) Antibodies (blocking polymerase activity) which denature at a lower temperature than the Polymerase enzyme.
http://www.abdserotec.com/anti-malaria-antibodies.html

(b) Properties of the enzyme itself – dependent on natural genetic variation and/or protein engineering.
http://www.sciencedirect.com/science/article/pii/S0022286004009986

(c) Wax (or ice) which prevents substrates and enzymes mixing until the solid to liquid transition temperature.
http://openi.nlm.nih.gov/detailedresult.php?img=355888_i1536-2442-001-04-0001-f02&req=4

In the main text we focused on (c) but also (a) and (b) is being considered.

In the case of enzymatic reaction the race is between heat transfer and heat production. For the case of enzymes with high turnover rate, one can usually assume that reaction rates is limited by substrate and enzyme diffusion. There are enzymes with reaction rates faster than these diffusion limit. That is, faster that a so-called "catalytically perfect enzyme". This may be connected with the fact that with ER, we may move from a diffusion-controlled mode to a turbulency-controlled rate, wherein very large gradients exist at sub-µm scales and drive the two materials mixing. Catalyse seems to belong to this sub-class, but this may not be necessary or sufficient (see http://www.catalase.com/cataext.htm).

Catalase kinetics and thermodynamics is described in many books and sites
http://web.mit.edu/chrissu/Public/5310lab3.pdf
http://www.indiana.edu/~iubmtc/Biology/Bio5.html

Under optimal conditions, catalase turnover rate is $4 \times 10^8$ reactions/(sec·x molecule).
$2 H_2O_2 \rightarrow 2H_2O + O_2$ $\Delta H_o = -98.2$ kJ·mol$^{-1}$; $\Delta S = 70.5$ J/mol·K.
Heat produced = $\Delta G = \Delta H - T\Delta S$ = -98,200 - (T x 70.5) = -119 kJ/mole (at 300 K)
http://en.wikipedia.org/wiki/Hydrogen_peroxide

The heat transfer coefficient in water is $h$= 500 to 10,000 W/m$^2$K. However, it may be larger in the presence of turbulence. Only future experiments will show which we should use.
http://en.wikipedia.org/wiki/Heat_transfer_coefficient

Thermal conductivity of water = 0.67 W/mK
http://koolance.com/cooling101-heat-transfer
Free Convection - Water: 20 - 100 W/m$^2$K
http://www.engineeringtoolbox.com/convective-heat-transfer-d_430.html

Surface area of a 1 μm radius sphere = 4$\pi r^2$ = 12.6 x 10$^{-12}$ m$^2$
So heat transfer = 100 x 12.6 x 10$^{-12}$ x 300 = 3.6 x 10$^{-7}$ W

Catalase MW = 60 kDaltons (per monomer of a tetramer). We conservatively use a monomer, but in the future, a tetramer may be better (factor of 4).

Reaction rate = 10$^7$ moles per moles of enzyme (at 0.1 M H$_2$O$_2$)
A 10% catalase solution = 100 g /liter = 1.7 x 10$^{-3}$ M
Volume of 1 μm radius sphere = 4/3 $\pi r^3$ = 4.2 x 10$^{-15}$ liter
Moles of catalase per 1 μm radius sphere = 7.0 x 10$^{-18}$ moles
Heat produced = 10$^7$ s$^{-1}$ x 7.0 x10$^{-18}$ x 119 kJ = 8.0 x 10$^{-6}$ W
So the ratio of heat produced to heat transferred = 8 x 10$^{-6}$ W / 3.6 x 10$^{-7}$ W = 22

Thus, in agreement with initial observations, we find that self-sustained ER using catalyse is possible. Actually, our initial experiments suggests that bolus of catalase cocktail inside of bulk ice is self-sustaining. However, this is a fragile equilibrium depending on type of catalyse, geometry and composition of catalyse cocktail. Obviously, we need more experiments.

**Appendix 3: Using mass-spectroscopy to read-out ER-detector.**

The ER detector using catalyse will produce oxygen. Thus it could be spatially encoded by using three stable isotopes of oxygen: O$^{16}$, O$^{17}$ and O$^{18}$. This signal can be measured using ultra-sensitive mass spectrometry. The redundancy (three isotopes) can be used as a type of bar-coding to improve read-out signal/background or to obtain spatial resolution. Up to 10$^6$ voxels can be recognized. However, if we add some solid carbon to a mix, the CO$_2$ will be created and this opens the possibility of also using stable isotopes of carbon.

Let us now consider use of mass-spectrometers (MS) in detection of ER products. The natural compositions for carbon and oxygen are:

Carbon:     C$^{12}$ (98.9%), C$^{13}$ (1.1%)
Oxygen:     O$^{16}$ (99.76%), O$^{17}$ (0.038%), O$^{18}$ (0.2%).

C$^{13}$, O$^{17}$ and O$^{18}$ are readily available and relatively cheap. They can be used in synthesis of selected explosives and in each particular batch of isotope enriched water, the ER can be tagged by four parameters. If gas after ER is taken into a mass-spectrometer, we can observe five (CO$_2$, O$^{16}$, O$^{17}$, O$^{18}$) peaks with very well distinguished mass. Actually, also C and O in CO$_2$ may be enriched in heavier isotopes, leading not to one, but three peaks. Furthermore as oxygen is escaping as O$_2$ molecules, we could identify O$^{17}$=O$^{17}$ *vs.* O$^{17}$=O$^{18}$ *vs.* O$^{18}$=O$^{18}$.

When using MS, the measurement of relative abundance of stable isotopes is limited by statistical uncertainties, *i.e.* the measurement has a precision of ($\Delta n$)/n w molecules in the peak, one can recognize 100 x 100 x 100 x 100 = 10$^8$ combinations, *i.e.* places inside the detector ordered by any combination of 4-isotopes. Practically, the limits is not the detection of stable isotope bar-codes but the ability to produce the bar-coded material. We believe, that 10$^6$ different voxels can be distinguished by their bar-code. We do not need to use MS in a continuous mode, *e.g.*, it can be triggered by the sonic boom localization. Having these two *independent* read-out methods will permit extra reliable detection with very high spatial resolution.

*Appendix 4: Energy Loss of Background:*

We previously showed that the Lindhard equation is used to calculate dE/dx for a slowly moving heavy ion moving through a medium (typical energies for the ion are a few keV): this equation is relevant for recoiling nuclei due to WIMP interactions.

In contrast, the Bethe-Bloch equation is used for relativistic heavily ionizing particles like alphas and betas. Each equation is valid only in specific considerations. More accurate expressions for the Lindhard Equation [A4.3] and the Bethe-Bloch Equation [A4.4] are readily found in literature and used for precise experiments and calculations. The analytical expressions are not only constrained to given assumptions but are also difficult to implement for most practical purposes. Fortunately, large tables have been experimentally collected and sophisticated simulations constructed in order to predict the stopping power of any type of particle. In particular, the stopping power for betas and alphas are found easily with ESTAR and ASTAR respectively [A4.5,A4.6]. Heavy ions are best simulated using the software TRIM [A4.2]. These three programs give the stopping range of a given beta particle, alpha particle or heavy ion with a particular energy moving through a specified medium. The stopping power for beta and alpha particles are important to consider because they are a source of background in most dark matter detectors. Understanding their energy deposition and range may elucidate ways to distinguish backgrounds from a WIMP signal. Another source of background are gammas, which are highly energetic photons.

Even though gamma particles are electrically neutral, they can ionize atoms directly through the photoelectric effect and the Compton effect. Either of those interactions will eject an electron at relativistic speeds, turning it into a beta particle that will ionize many more atoms. Typical energies for gamma particles produced in a radioactive decay range from a few hundred keV to 10MeV. ESTAR can be employed to learn the behavior of the electrons produced by the interaction of an atom with a gamma ray.

Utilizing TRIM, ESTAR and ASTAR one can characterize the typical range of an ion or background (alpha, beta or gamma particles) moving through a specified medium. The usual range for heavy ions is tens of nanometers; whereas alpha, beta and gamma particles have a range of several hundred nanometers or greater. The range of particles is dependent on the medium, the type of particle and its initial energy. Thus, a detector with spatial resolution of a few tens of nanometers would be able to distinguish between background and true signals; assuming that true signals come from a WIMP interacting with a nucleus and producing a recoiled ion. Thus, signals can be differentiated from background: the recoiling nuclei from WIMP interactions stop within tens of nanometers while the background particles travel much further, at least several hundreds of nanometers.

**References to Appendix 4:**

[A4.1] Yu, L. D., T. Kamwanna, and I. G. Brown. "The Low-energy Ion Range in DNA."*Physics in Medicine and Biology* 54.16 (2009): 5009-022. Print
[A4.2] J.F. Zigler, Program SRIM/TRIM, version 2013, obtained from http://www.srim.org.
[A4.3] I.S. Tilinin, Phys. Rev. A 51 (1995) 3058.
[A4.4] Steven P. Ahlen, Erratum: Theoretical and experimental aspects of the energy loss of relativistic heavily ionizing particles, Rev. Mod. Phys. 52, 653 (1980)
[A4.5] M.J. Berger, J.S. Coursey, M.A. Zucker and J. Chang, database ESTAR, obtained from http://www.nist.gov/pml/data/star/index.cfm
[A4.6] M.J. Berger, J.S. Coursey, M.A. Zucker and J. Chang, database ASTAR, obtained from http://www.nist.gov/pml/data/star/index.cfm